\documentclass{amsart}
\usepackage{amsmath,amssymb,amsthm}
\usepackage{amsfonts}
\usepackage{amscd}
\usepackage[ansinew]{inputenc}
\usepackage{verbatim}
\usepackage{graphicx}

\newcommand{\be}{\begin{equation}}
\newcommand{\ee}{\end{equation}}
\newcommand{\bea}{\begin{array}}
\newcommand{\ea}{\end{array}}
\newcommand{\beqa}{\begin{eqnarray}}
\newcommand{\eeqa}{\end{eqnarray}}
\newcommand{\bean}{\begin{eqnarray*}}
\newcommand{\eean}{\end{eqnarray*}}

\theoremstyle{definition}
\newtheorem{Def}{Definition}
\newcommand{\gapproxeq}{\lower
 .7ex\hbox{$\;\stackrel{\textstyle >}{\sim}\;$}}
\newcommand{\lapproxeq}{\lower .7ex\hbox{$\;\stackrel
{\textstyle <}{\sim}\;$}}

\def\up#1{\leavevmode \raise.16ex\hbox{#1}}

\newcounter{appendice}

\def\thebibliography#1{{\bf REFERENCES\markboth
 {REFERENCES}{REFERENCES}}\list
 {[\arabic{enumi}]}{\settowidth\labelwidth{[#1]}\leftmargin\labelwidth
 \advance\leftmargin\labelsep
 \usecounter{enumi}}
 \def\newblock{\hskip .11em plus .33em minus -.07em}
 \sloppy
 \sfcode`\.=1000\relax}

\begin{document}

%\vspace*{5mm}
\centerline{\Large Noncommutative Coherent States and Quantum Cosmology}
\vskip .5cm
\centerline{ {Rom\'an Ju\'arez$^{a}$} and {David Mart\'inez$^{b}$} }
\vskip 1cm

\begin{center}
Instituto de Ciencias Nucleares,\\
Universidad Nacional Aut\'onoma de M\'exico,\\
 A. Postal 70-543 , M\'exico D.F., M\'exico \\
{\it a)roman.juarez@nucleares.unam.mx}
\end{center}
\begin{center}
Instituto de Investigaci\'on en Matem\'aticas Aplicadas y en Sistemas,\\
Universidad Nacional Aut\'onoma de M\'exico,\\
{\it b)dmr@mym.iimas.unam.mx}
\end{center}

\date{\today}

\vskip 2cm \vspace*{5mm} \normalsize \centerline{\bf ABSTRACT}

The set of coherent states for a noncommutative quantum Bianchi I anisotropic cosmology were built to circumvent the absence of a simultaneous set of configuration observables. By extending known methods of path integrals with coherent states to their noncommutative analogues allowed us to obtain the formal expression of the propagator of the theory in terms of the covariant Husimi Q-symbol of the quantum constraint. The analysis of the equations of motion resulting from a steepest descent procedure showed the existence of solutions displaying a minimum value of the volume scaling function which are, in turn, compatible with a physically inspired definition for a bounce. More importantly, a lower bound for the volume at the bounce which incorporates quantum mechanical and noncommutativite contributions was established. The asymptotic analysis of the solutions in the vicinity of the bounce was performed by implementing techniques used in boundary-layer problems. The numerical simulations that confirm our results are also presented.

{\bf PACS: 03.70.+k, 98.80.Qc, 03.65.Sq}

{\bf 2010 MSC:} 81Q99, 81R60, 81S30, 81Q20, 81S40, 76M45
\keywords{Coherent States, Noncommutativity, Quantum Cosmology}

\newpage

\section{Introduction}
Since their introduction by Glauber \cite{Glau69}, coherent states have become fundamental objects in the study of a plethora of physical phenomena. Their properties have profound conceptual implications, such as the case of their non spreading nature which allows to read classical behaviour from quantum systems. As normalized supercomplete sets for Hilbert spaces they correctly encode all the probabilistic interpretation of Quantum Mechanics. In a more mathematical context they constitute the appropriate language to study harmonic analysis in Lie groups and, as it was shown in \cite{var90}, they bridge different deformation quantization schemes.
As for the latter it was also demonstrated in \cite{Ju10} that such bridge remains true even for the case of Noncommutative Quantum Mechanics understood as the quantum theory associated to the Lie algebra $\mathfrak{h}_{2n+1}^\theta$ of commutators:

\be\label{nchw}
[\hat Q^i,\hat Q^j]=i\theta^{ij}\mathbb{\hat I},\quad [\hat Q^i,\hat P_j]=i\hbar\delta^i_j\mathbb{\hat I},\quad [\hat P_i,\hat P_j]=0, \quad i,j=1,...,n,
\ee
where $\theta^{ij}$ is a real antisymmetric constant matrix.

The algebra (\ref{nchw}) can be justified by simultaneously applying the equivalence principle and the uncertainty principle \cite{Dop95}. According to which the measurement of the position of a particle in every spatial coordinate with ever higher precision leads to a minimum possible value, the Planck length $\ell_p=\sqrt{G\hbar\over c^3}$, beyond which smaller distances lack operational meaning due to the presence of horizons that prevent to extract information from such a region.\footnote{$\ell_p=1.6162\times 10^{-35}m$.} This implies an uncertainty principle among spatial coordinates $\Delta q^i\Delta q^j\geq\ell_p^2$ that in the language of Quantum Mechanics translates to the first commutator in (\ref{nchw}) if $|\theta^{ij}|\sim\ell_p^2$.

Research in Loop Quantum Cosmology viewed as the one parameter minisuperspace sector of Loop Quantum Gravity has shed light on the implications of implementing a smallest length scale as inherent property of the structure of space-time \cite{ash1}. There, the spectra of geometrical observables (volumes and areas) become discretized and the classical singularities are replaced by a bounce behaviour which connects collapsing solutions to expanding ones. This interesting aspect is actually the consequence of recurring to an appropriate Hilbert space for the theory known as the polymeric Hilbert space $\mathcal{H}_{poly}$.

Even more interesting is the fact that the usual normalizable basis of $\mathcal{H}_{poly}$ is closely related with the set of coherent states based on a regular lattice (\emph{cf.} Chap. 15 of \cite{gaz09}). In this direction it has also been shown that noncommutativity of space can dynamically reproduce features of field theories based on lattices by studying the evolution of gaussian (coherent) states \cite{Min09}. Therefore, it is tempting to implement the full coherent-state formalism in the case of a Noncommutative Quantum Cosmology, where the configuration observables will be described by the algebra (\ref{nchw}), in the hope that at an effective level the dynamical properties of noncommutativity suffice to generate solutions which are free of singularities. Overcomplete sets of coherent states have been constructed elsewhere for the case of noncommutative theories \cite{Ju10,Smai03,scholtz09,gazeau09}, where it must be emphasized that the representations used in all of these references are not equivalent and consequently the results obtained from their implementation in a given problem should differ. In the present work, however, we will opt for the use of the coherent system developed in \cite{Ju10}.

Thus, to investigate the consequences of introducing a noncommutative structure at the level of a Quantum Cosmology we organized this work in the following manner: In \S2 and \S3 we review the basics of Perelomov's definition of Generalized Coherent States and reproducing kernels \cite{pere72, pere86}, with their application to Feynman integrals of constraint theories by means of projectors, according to Klauder's procedure of coherent state quantization \cite{Klau97}. We then devote \S4 and \S5 to construct the generalizations of these concepts for Noncommutative Quantum Mechanics, where the operators representing position observables satisfy non trivial commutators. In \S6 we briefly summarize the anisotropic Bianchi I model in the ADM separation by selecting the scale factors from the metric and their conjugate momenta as dynamical variables. Then in \S7, by promoting these variables to quantum operators that satisfy a noncommutative algebra, we give a notion of a Noncommutative Quantum Cosmology. Here, we construct the quantum equivalent of the Hamiltonian constraint in terms of ladder operators, associated to the observables of the theory, and we obtain the corresponding coherent state Husimi Q-symbol in the path integral action. In \S8 we analyze the equations of motion from a stationary phase evaluation of the action. A bounce of the system is defined in \S9 and its implications on the solutions are studied. In \S10 we provide various numerical solutions that illustrate the bounce-like behavior at Planckian scales. Finally in \S11 we discuss our results and present our conclusions.

\section{Generalized Coherent States}

The existing definitions for a coherent state are just as numerous as its applications (see \emph{e.g.}, \cite{Gil90, gaz90, gaz09} and references therein). Perelomov's definition of generalized coherent state \cite{pere72, pere86} is adequate to study physical systems with an underlying symmetry group, as is the case of \emph{elementary systems} \cite{var90}.

\begin{Def}[Coherent State]
An elementary (classical) system corresponds to a symplectic manifold $X$ which is homogeneous under the action of a Lie group $G$ and, in consequence, isomorphic to the orbits generated by a maximal subgroup (the isotropy group) $H\subset G$, meaning $X\simeq G/H$. If in addition $G$ is semisimple and acts on a Hilbert space via some UIR $\hat U(g)$ and $H$ is compact, then, the coherent states of $\mathcal{H}$ tagged by points of $X$ are the vectors obtained via the transitive action of $G$ on the $H$ invariant subspace $\mathcal{H}_0\subset\mathcal{H}$.
\end{Def}

Thus the first step in building the CS system is picking a fiducial normalized state $|\varphi_0\rangle\in\mathcal{H}_0$, called the ground state, which due to the $H$-invariance satisfies
\be
\hat U(h)|\varphi_0\rangle=e^{i\rho(h)}|\varphi_0\rangle,\; h\in H,
\ee
where $\rho(h)$ is a real valued function of $h$.

Now, because $G$ admits a decomposition in terms of left cosets, \emph{i.e.} $G=\{g_xh|\;g_x\in G/H,\, h\in H\}$, the action of an arbitrary element $g\in G$ over $|\varphi_0\rangle$ is given by
\be\label{gaction}
\hat U(g)|\varphi_0\rangle=\hat U(g_x)\hat U(h)|\varphi_0\rangle=e^{i\rho(h)}\hat U(g_x)|\varphi_0\rangle,\quad\forall g\in G,
\ee
where $g_x$ is identified with the geometric point $x\in X$. In this way the vector $\hat U(g_x)|\varphi_0\rangle$ from the previous expression is, according to the definition, the \emph{generalized coherent state} associated to the point $x$:
\be\label{gcs}
|x\rangle:=\hat U(g_x)|\varphi_0\rangle,\quad\forall g_x\in G/H,
\ee
which, by construction, will also be a normalized state of $\mathcal{H}$.

The first and foremost property of the set $\{|x\rangle\}$ is that, due to Schur's lemmas, there is a measure $d\mu(x)$ for which the coherent states form a complete basis of $\mathcal{H}$, in other words, there's a resolution of unity:
\be\label{iresolution}
\int_Xd\mu(x)\;|x\rangle\langle x|=\mathbb{\hat I}_\mathcal{H},
\ee
where $d\mu(x)$ is actually the Riemann invariant measure in $X$.

The second property of $\{|x\rangle\}$ is the modulus of the transition function between any two coherent states:
\be
0<|\langle x|x'\rangle|\leq1,
\ee
which, along with the resolution of unity, implies the well known fact that the coherent states constitute an over complete basis.

The bounded and continuous function $K(x',x):=\langle x'|x\rangle$ provides an example of a reproducing kernel \cite{aron50}, meaning
\be\label{kernel}
K(x',x)=\int_Xd\mu(y)K(x',y)K(y,x),\;\forall x',x,
\ee
an identity that may appear trivial simply by making use of (\ref{iresolution}). However, contrary to what happens in the case of Dirac-$\delta$ functions, by setting $x'=x$ and noticing $K(x,y)=K^*(y,x)$ it is then clear that $K_y:=K(y,\cdot)\in\mathcal{H}$.

Also by evaluating any state in the coherent basis $\langle x|\psi\rangle=\psi(x)$ it is seen that
\be\label{density}
\psi(x)=\int_Xd\mu(y)K(x,y)\psi(y),
\ee
which expresses the non-local nature of $K$ (the integral has actually to be carried out).

In what follows we will use the above definition of coherent state to obtain the noncommutative equivalent of the path integral formalism, but first we will review the necessity of a well defined reproducing kernel for the case of constraint theories.

\section{Constrained Reproducing Kernel}

When working with the canonical quantization scheme for a constraint system \cite{Dir64}, once all the second--class constraints have been removed, the supplementary quantization condition that unambiguously ensures the evolution of a physical quantum state demands:
\be\label{constraint}
\hat\Phi_a|\psi_{phys}\rangle=0, \forall\hat\Phi_a,
\ee
where $\{\hat\Phi_a\}$ is the set of first--class quantum constraints.

Expression (\ref{constraint}) establishes the appropriate Hilbert space $\mathcal{H}_{phys}$ for the theory. Thus if $\mathcal{H}$ constitutes a Hilbert space for a suitable representation of the operators that characterize the system, known as the kinematical Hilbert space, then, only the vector subspace $\mathcal{H}_{phys}:=\{|\psi\rangle\in\mathcal{H}|\;\hat\Phi_a|\psi\rangle=0,\forall\Phi_a\}$ will be of main interest.

If we decide to use a basis of coherent states for $\mathcal{H}$ then there will be an overcomplete subset of $\{|x\rangle\}$ contained in $\mathcal{H}_{phys}$. This basis and its associated kernel shall be obtained by means of a projector which can be constructed using the method of group averaging with the constraints \cite{Klau97}:
\be\label{projector}
\mathbb{\hat E}:=\int d\mu(\sigma)e^{-i\sum_a\sigma_a\hat\Phi_a},
\ee
where $d\mu(\sigma)$ is a (normalized) invariant measure on the group manifold generated by the algebra of constraints.\footnote{Observe that the $\sigma_a$ must have the appropriate units so the argument in the exponential is dimensionless.} Due to the measure invariance and the Lie algebra of the constraints, $\mathbb{\hat E}$ is a self-adjoint and idempotent operator:
\be
\mathbb{\hat E}^\dag=\mathbb{\hat E},\quad\mathbb{\hat E}^2=\mathbb{\hat E},
\ee
so (\ref{projector}) defines a projector indeed.

The operator $\mathbb{\hat E}$ produces a surjective endomorphism $\mathcal{H}\twoheadrightarrow\mathcal{H}_{phys}$ which now may be implemented to obtain the corresponding reproducing kernel in $\mathcal{H}_{phys}$. By noticing that
\be\label{kernel2}
\langle x'|\mathbb{\hat E}|x\rangle=\langle x'|\mathbb{\hat E}^2|x\rangle=\int_Xd\mu(y)\langle x'|\mathbb{\hat E}|y\rangle\langle y|\mathbb{\hat E}|x\rangle,
\ee
and after comparing with (\ref{kernel}) it shows that the projected transition function $K^\Phi(x',x):=\langle x'|\mathbb{\hat E}|x\rangle$ is a reproducing kernel too.

Therefore, similarly to equation (\ref{density}), when projected on the basis $\{|x\rangle\}$, any state $|\psi_{phys}\rangle$ will have the form
\be
\psi_{phys}(x)=\int_Xd\mu(y)K^\Phi(x,y)\psi_{phys}(y)
\ee
which together with (\ref{kernel2}) are expressions purely in $\mathcal{H}_{phys}$ as intended.

\section{Noncommutative Coherent States of $H_{2n+1}^\theta$}
By non-commutative coherent states we will mean the states defined in expression (\ref{gcs}) where the corresponding symmetry group $G$ is the extended Heisenberg-Weyl group $H_{2n+1}^\theta$ whose Lie algebra $\mathfrak{h}_{2n+1}^\theta$ generators satisfy the commutators (\ref{nchw}).

Thus $\mathfrak{h}_{2n+1}^\theta$ corresponds to the algebra of Quantum Mechanics, but where position observables do not commute anymore, implying a new uncertainty principle that sets a minimum value in the precision of any attempt to locate a particle in space:
\be
\label{uncert1}\Delta Q_i\Delta Q_j\geq{|\theta_{ij}|\over2}.
\ee

In this case a simultaneous basis for position operators is no longer possible, and this will represent our main motivation behind using coherent states as it should be evident in what follows. It is worth mentioning that other bases are also admissible (see \emph{e.g.} \cite{Ros08}), for example of eigenstates of $\hat P$ operators, or one spatial coordinate and $n-1$ momenta, although contrary to what happens with a coherent state basis none of those belong to $\mathcal{H}$.

The typical element of $H_{2n+1}^\theta$ is the exponentiation $g(c,\underline{q},\underline{p})=e^{\chi(c,\underline{q},\underline{p})}$ where $\chi(c,\underline{q},\underline{p})\in\mathfrak{h}_{2n+1}^\theta$ corresponds to
\be
\chi(c,\underline{q},\underline{p})=ic\mathbb{\hat I}+{i\over\hbar}\sum_{i=1}^n(p_i\hat{Q}_i-q_i\hat{P}_i);\quad c\in\mathbb{R},\; \underline{q}\in\mathbb{R}^n,\; \underline{p}\in\mathbb{R}^n.
\ee
If we now make use of alternate generators ({\it cf.} \cite{Ju10})
\be\label{ncanni}
\begin{split}
\hat A_i:={1\over\sqrt{2\hbar}}(\hat Q_i+{1\over2\hbar}\sum_{j=1}^n\theta_{ij}\hat P_j+i\hat P_i),\\
\hat A_i^\dag:={1\over\sqrt{2\hbar}}(\hat Q_i+{1\over2\hbar}\sum_{j=1}^n\theta_{ij}\hat P_j-i\hat P_i),
\end{split}
\ee
the original position and momenta operators can be written as
\be
\begin{split}
\hat Q_i&={1\over\sqrt{2\hbar}}\bigg[\hat A_i+\hat A_i^\dag+{i\over2\hbar}\sum_{j=1}^n\theta_{ij}(\hat A_j-\hat A_j^\dag)\bigg],\\
\hat P_i&={1\over i\sqrt{2\hbar}}(\hat A_i-\hat A_i^\dag).
\end{split}
\ee
The group element $g(c,\underline{q},\underline{p})$ in terms of the new generators is
\be\label{gele}
g(c,\underline{q},\underline{p})=g(c,\underline{\alpha},\underline{\alpha}^*)=e^{ic\mathbb{\hat I}}\exp\bigg[\sum_{i=1}^n(\alpha_i\hat A_i^\dag-\alpha_i^*\hat A_i)\bigg],
\ee
with the parameters $\alpha_i,\alpha_i^*\in\mathbb{C}$.

Operators $\hat A_i, \hat A_i^\dag$ satisfy the $n$-dimensional creation--annihilation commutator algebra
\be\label{caalg}
[\hat A_i,\hat A_j^\dagger]=\delta_{ij},\quad[\hat A_i,\hat A_j]=[\hat A_i^\dagger,\hat A_j^\dagger]=0.
\ee
where each pair $\hat A_i, \hat A_i^\dag$ spans a Hilbert space $\mathcal{H}_i$ in the Fock basis $\{|m_i\rangle\}$, satisfying
\be
\begin{split}
\hat A_i|m_i\rangle&=\sqrt{m_i}|m_i-1\rangle,\\
\hat A_i^\dag|m_i\rangle&=\sqrt{m_i+1}|m_i+1\rangle,\\
|m_i\rangle&={(A_i^\dag)^{m_i}\over\sqrt{m_i!}}|0\rangle,
\end{split}
\ee
so the entire Hilbert space $\mathcal{H}=\bigotimes_{i=1}^n\mathcal{H}_i$ is spanned by the basis $\{|m_1,...,m_n\rangle\}$

From (\ref{gele}) it is clear that $\mathbb{\hat I}$ generates the maximal compact subgroup $U(1)$ which trivially leads to the homogeneous space $\mathbb{C}^n=H_{2n+1}^\theta/U(1)$. Thus according to expressions (\ref{gaction}) and (\ref{gcs}) from the general theory, the noncommutative coherent states of $\mathcal{H}$ tagged by points of $\mathbb{C}^n$ are those obtained from the action
\be\label{nccs}
|\underline{\alpha}\rangle=\exp\bigg[\sum_{i=1}^n(\alpha_i\hat A_i^\dag-\alpha_i^*\hat A_i)\bigg]|0\rangle=\prod_{i=1}^ne^{(\alpha_i\hat A_i^\dag-\alpha_i^*\hat A_i)}|0\rangle,
\ee
which are just the usual Glauber-Sudarshan  coherent states corresponding to the eigenstates $\hat A_i|\underline{\alpha}\rangle=\alpha_i|\underline{\alpha}\rangle$, with $\hat A_i$ given by (\ref{ncanni}).

An important consequence, due to the these considerations, is that the set of phase-space variables $\{q_i,p_i,\;i=1,...,n\}$ are precisely the expectations
\be\label{exval}
\begin{split}
\langle\underline{\alpha}|\hat Q_i|\underline{\alpha}\rangle&=q_i={1\over\sqrt{2\hbar}}\bigg[\alpha_i+\alpha_i^*+{i\over2\hbar}\sum_{j=1}^n\theta_{ij}(\alpha_j-\alpha_j^*)\bigg],\\
\langle\underline{\alpha}|\hat P_i|\underline{\alpha}\rangle&=p_i={1\over i\sqrt{2\hbar}}(\alpha_i-\alpha_i^*),\\
\end{split}
\ee
and so they are quantities that can be specified simultaneously and do not constitute a major conflict, contrasting with the case of eigenvalues of $\hat Q_i$.

The resolution of unity is given by
\be\label{ncunity}
\begin{split}
&\mathbb{\hat I}_\mathcal{H}=\int_{\mathbb{C}^n}d\mu(\underline{\alpha},\underline{\alpha}^*)|\underline{\alpha}\rangle\langle\underline{\alpha}|,\\
&d\mu(\underline{\alpha},\underline{\alpha}^*)={1\over(2\pi i)^n}\prod_{i=1}^nd\alpha_id\alpha_i^*,
\end{split}
\ee
and the reproducing kernel corresponds to
\be\label{ncrepkernel}
K(\underline{\alpha}^*,\underline{\beta})=\langle\underline{\alpha}|\underline{\beta}\rangle=e^{\left(-{1\over2}|\underline{\alpha}|^2-{1\over2}|\underline{\beta}|^2+\underline{\alpha}^*\cdot\underline{\beta}\right)}.
\ee

These results show that through definitions (\ref{ncanni}) one recovers the well known properties of the bosonic quantum mechanical coherent states with no explicit presence of the noncommutativity of the spatial coordinates, as long as one remains in the holomorphic variables. This property has been studied in the context of star products in \cite{Ju10}.

%From a purely algebraic perspective it might look like there's a Bopp shift underlying, which usually appears in most of the literature related to noncommutativity, however, no assumptions have been made here about observables and their eigenvalues.
We are thus in position to derive the corresponding path integral, by taking advantage of the fact that at the level of holomorphic coordinates the construction should not differ from the approach in \cite{Klau97}, and only when going back to physical variables the differences will become clear.

\section{Noncommutative Coherent State Path Integral}

Following the construction of the path integral for constraint theories in terms of Coherent States developed in \cite{Klau97}, and, for clarity sake, we will proceed with a similar calculation in the case of the coherent state system (\ref{nccs}). Furthermore we will specialize to the case of a purely constrained system with a single constraint, \emph{i.e.}, when the total Hamiltonian is only the constraint $\hat\Phi$.

As it was discussed in a previous section, the adequate reproducing kernel for the present case will be given by
\be\label{nckernel}
\langle\underline{\alpha}''|\mathbb{\hat E}|\underline{\alpha}'\rangle=\int d\mu(\sigma)\langle\underline{\alpha}''|e^{-i\sigma\hat\Phi}|\underline{\alpha}'\rangle.
\ee
Because there's no actual Hamiltonian, there is no evolution and the reproducing kernel will effectively act as the propagator. This can be shown by trying to evolve an arbitrary physical state $ |\psi_{phys}\rangle=\mathbb{\hat E}|\psi\rangle$ with the constraint
\be
|\psi_{phys},\tau\rangle=e^{-i\tau\hat\Phi}|\psi_{phys}\rangle=e^{-i\tau\hat\Phi}\mathbb{\hat E}|\psi\rangle,
\ee
and observing that
\be
e^{-i\tau\hat\Phi}\mathbb{\hat E}=\int d\mu(\sigma)e^{-i(\sigma+\tau)\hat\Phi}=\mathbb{\hat E},
\ee
we obtain
\be
|\psi_{phys},\tau\rangle=|\psi_{phys}\rangle,
\ee
which implies that the physical states are frozen and the transition amplitude of going from an initial physical state to a final physical state will depend only on the reproducing kernel (\ref{nckernel}).$\qed$

To evaluate the reproducing kernel by the Feynman procedure of infinitesimal time slices we will need to introduce a fictitious interval $N\epsilon=1$:
\be\label{pathint1}
\begin{split}
\langle\underline{\alpha}''|\mathbb{\hat E}|\underline{\alpha}'\rangle&=\langle\underline{\alpha}''|\int d\mu(\sigma)e^{-i\sigma\hat{\Phi}}|\underline{\alpha}'\rangle\\
&=\int d\mu(\sigma)\langle\underline{\alpha}''|\underbrace{e^{-i\epsilon\sigma\hat{\Phi}}\times...\times e^{-i\epsilon\sigma\hat{\Phi}}}_N|\underline{\alpha}'\rangle\\
&=\int d\mu(\sigma)\int\prod_{k=0}^{N-1}d\mu(\underline{\alpha}_k,\underline{\alpha}_k^*)\langle\underline{\alpha}_{k+1}|e^{-i\epsilon\sigma\hat{\Phi}}|\underline{\alpha}_k\rangle,\\
\end{split}
\ee
after making use of (\ref{ncunity}) and where $\underline{\alpha}_N=\underline{\alpha}''$ and $\underline{\alpha}_0=\underline{\alpha}'$.

If $\hat\Phi$ is already a normal ordered operator, meaning it can be  formally written as
\be\label{normalorder}
\hat\Phi=\prod_{i=1}^n\sum_{r_i,s_i}\xi_{r_i,s_i}(\hat{A}_i^\dag)^{r_i}(\hat{A}_i)^{s_i},
\ee
then by focusing on the $k$-th term in expression (\ref{pathint1}) at leading order $\epsilon$ we have
\be
\begin{split}
\langle\underline{\alpha}_{k+1}|e^{-i\epsilon\sigma\hat{\Phi}}|\underline{\alpha}_k\rangle&\approx\langle\underline{\alpha}_{k+1}|\underline{\alpha}_k\rangle-i\epsilon\sigma\langle\underline{\alpha}_{k+1}|\hat{\Phi}|\underline{\alpha}_k\rangle\\
&=\langle\underline{\alpha}_{k+1}|\underline{\alpha}_k\rangle[1-i\epsilon\sigma\Phi(\underline{\alpha}_{k+1}^*,\underline{\alpha}_k)]\\
&\approx\langle\underline{\alpha}_{k+1}|\underline{\alpha}_k\rangle e^{-i\epsilon\sigma\Phi(\underline{\alpha}_{k+1}^*,\underline{\alpha}_k)},
\end{split}
\ee
with
\be
\Phi(\underline{\alpha}_{k+1}^*,\underline{\alpha}_k)=\prod_{i=1}^n\sum_{r_i,s_i}\xi_{r_i,s_i}(\alpha_{\{k+1\}i}^*)^{r_i}(\alpha_{\{k\}i})^{s_i},
\ee
where notation $\alpha_{\{k\}i}$ implies the $i$-th component of vector $\underline{\alpha}_k$.

Thus we have the approximation
\be\label{pathint2}
\langle\underline{\alpha}''|\mathbb{\hat E}|\underline{\alpha}'\rangle\approx\int d\mu(\sigma)\int\prod_{k=0}^{N-1}d\mu(\underline{\alpha}_k,\underline{\alpha}_k^*)\langle\underline{\alpha}_{k+1}|\underline{\alpha}_k\rangle e^{-i\epsilon\sigma\Phi(\underline{\alpha}_{k+1}^*,\underline{\alpha}_k)},
\ee
and lastly by using (\ref{ncrepkernel}) we have
\be
\begin{split}
\langle\underline{\alpha}_{k+1}|\underline{\alpha}_k\rangle&=\exp\left[-{\underline{\alpha}_{k+1}^*\cdot\underline{\alpha}_{k+1}\over 2}-{\underline{\alpha}_{k}^*\cdot\underline{\alpha}_{k}\over 2}+\underline{\alpha}_{k+1}^*\cdot\underline{\alpha}_{k}\right]\\
&=\exp\left[\epsilon\left({\underline{\alpha}_{k}\over 2}\cdot{(\underline{\alpha}_{k+1}^*-\underline{\alpha}_{k}^*)\over\epsilon}-{\underline{\alpha}_{k+1}^*\over 2}\cdot{(\underline{\alpha}_{k+1}-\underline{\alpha}_{k})\over\epsilon}\right)\right],
\end{split}
\ee
which when substituted in (\ref{pathint2}) gives
\be
\begin{split}
\langle\underline{\alpha}''|\mathbb{\hat E}|\underline{\alpha}'\rangle\approx\int &d\mu(\sigma)\int\prod_{k=0}^{N-1}d\mu(\underline{\alpha}_k,\underline{\alpha}_k^*)\bigg\{e^{-i\epsilon\sigma\Phi(\underline{\alpha}_{k+1}^*,\underline{\alpha}_k)}\\
&\times\exp\left[\epsilon\left({\underline{\alpha}_{k}\over 2}\cdot{(\underline{\alpha}_{k+1}^*-\underline{\alpha}_{k}^*)\over\epsilon}-{\underline{\alpha}_{k+1}^*\over 2}\cdot{(\underline{\alpha}_{k+1}-\underline{\alpha}_{k})\over\epsilon}\right)\right]\bigg\},
\end{split}
\ee
that in the limit $N\rightarrow\infty$ becomes the exact expression
\be\label{pathintf}
\begin{split}
\langle\underline{\alpha}''|\mathbb{\hat E}|\underline{\alpha}'\rangle=\int_{\underline{\alpha}'}^{\underline{\alpha}''}\mathcal{D}\mu(\underline{\alpha},\underline{\alpha}^*)\int d\mu(\sigma)e^{i\int_0^1 dt[{i\over2}(\underline{\alpha}^*(t)\cdot\underline{\dot\alpha}(t)
-\underline{\alpha}(t)\cdot\underline{\dot\alpha}^*(t))-\sigma\Phi(\underline{\alpha}(t),\underline{\alpha}^*(t))]}.
\end{split}
\ee
This form for the path integral confirms the properties discussed for coherent states (\ref{nccs}), in that there is no explicit appearance of the noncommutativity. To recover the noncommutative version of the path integral we make use of expressions (\ref{exval}) to replace the holomorphic coordinates in (\ref{pathintf}) by phase-space variables. It can be easily checked that the result is
\be\label{pathintfNC}
\begin{split}
\langle{\underline q''},{\underline p''}|\mathbb{\hat E}|{\underline q'},{\underline p'}\rangle=\int_{{\underline q'},{\underline p'}}^{{\underline q''},{\underline p''}}\mathcal{D}\mu({\underline q},{\underline p})\int d\mu(\sigma)e^{i\int_0^1 dt\big[{1\over\hbar}\sum_{i}\big(p_i{\dot q_i}+\sum_j{\theta_{ij}\over2\hbar}p_i{\dot p_j}\big)-\sigma\Phi({\underline q}(t),{\underline p}(t))\big]},
\end{split}
\ee
from which the noncommutative action associated to this path integral reads
\be\label{NCaction}
S=\int_0^1 dt\big[\sum_{i}\big(p_i{\dot q_i}+\sum_j{\theta_{ij}\over2\hbar}p_i{\dot p_j}\big)-\tilde\sigma\Phi({\underline q}(t),{\underline p}(t))\big],\; \tilde\sigma=\hbar\sigma.
\ee

Formula (\ref{pathintfNC}) for the noncommutative path integral coincides (in its functional form, although inequivalent) with similar versions obtained via different approaches, see, \emph{e.g}, \cite{Acat01,Ros07}, however, because this representation was obtained by making use of the coherent state kernel (\ref{nckernel}) it has the advantages detailed in \cite{Klau97}. In particular, it is gauge invariant and because the associated Hamiltonian is the covariant Husimi Q-symbol \cite{hus40} of the quantum constraint, one can expect a stationary phase evaluation (along classical trajectories) to be a fairly good approximation \cite{Bar01}.

\section{Classical Anisotropic Bianchi I cosmology}

In order to be able to introduce a noncommutative relation between the space-like field observables of the quantum theory, of the type discussed in the Introduction and section 4, it is necessary to work with a system of more than one spatial coordinate. Therefore as a classical model we will recur to the anisotropic Bianchi I cosmology, defined by the line element

\be\label{bianchi}
ds^2=-N^2(\tau)d\tau^2+g_{ij}(\tau)dx^idx^j,\quad g_{ij}(\tau)=a_i^2(\tau)\delta_{ij},
\ee
where $N(\tau)$ is the lapse function and the dimensionless quantities $a_i(\tau)$ are the scales of the cosmology in three independent space-like directions.

Because the $3$-curvature tensor vanishes, the corresponding Einstein-Hilbert action written in terms of the ADM canonical separation corresponds to
\be\label{classact}
\begin{split}
S&={\kappa}\int d\tau d^3x\big[\pi^{ij}\dot g_{ij}-{N(\tau)\over\sqrt{^{(3)}g}}\big(\pi^{ij}\pi_{ij}-{1\over2}(\pi^i_i)^2\big)\big]\\
&={\kappa}\int d\tau d^3x\big[\pi^{ij}\dot g_{ij}-{N(\tau)\over\sqrt{^{(3)}g}}\big(\pi^{ij}\pi^{kl}g_{ik}g_{jl}-{1\over2}(\pi^{ij}g_{ij})^2\big)\big],
\end{split}
\ee
with $^{(3)}g=Det(g_{ij})$ and $\pi^{ij}$ being the momenta conjugate to $g_{ij}$.\footnote{$\kappa={c^3/G}$. For later convenience during our quantization prescription we decide to retain all the units.}

Substituting the components $g_{ij}$ explicitly in the action above and by using the canonical transformation $\pi^i:=2\pi^{ij}a_j$, we can recast (\ref{classact}) in terms of the canonical pair $(a_i,\pi^i)$ as
\be\label{classact2}
S=\kappa\int d\tau d^3x\big[\pi^i\dot a_i-{N(\tau)\over4\sqrt{^{(3)}g}}\big((\pi^i)^2(a_i)^2-{1\over2}(\pi^ia_i)^2\big)\big],
\ee
with the Hamiltonian constraint
\be\label{classconst}
\mathcal{C}_{grav}={N(\tau)\over4\sqrt{^{(3)}g}}\big((\pi^i)^2(a_i)^2-{1\over2}(\pi^ia_i)^2\big).
\ee

The equations of motion may now be obtained by fixing ${N(\tau)\over4\sqrt{^{(3)}g}}=1$ and taking the variations $\delta a_i$ and $\delta\pi^i$:
\be\label{classdyn}
\begin{split}
\dot{a}_i&=2\delta^j_i\pi^ja_j^2-(\pi^ja_j)a_i=[2\delta^j_i\pi^ja_j-(\pi^ja_j)]a_i,\\
\dot{\pi}^i&=-2\delta^i_j(\pi^j)^2a_j+\pi^i(\pi^ja_j)=-[2\delta^i_j\pi^ja_j-(\pi^ja_j)]\pi^i,
\end{split}
\ee
and because the quantities $\pi^1a_1,\pi^2a_2,\pi^3a_3$ are clearly constants of motion, the solutions are of the form
\be\label{classsol}
\begin{split}
a_i(\tau)&=a_i(\tau_0)e^{(\tau-\tau_0)\eta_i},\\
\pi^i(\tau)&=\pi^i(\tau_0)e^{-(\tau-\tau_0)\eta_i},
\end{split}
\ee
where $\eta_i=(2\pi^ia_i-\pi^ja_j)|_{\tau_0}$, with no sum over $i$.

Depending on the sign of $\eta_i$, solutions (\ref{classsol}) imply the well known asymptotic collapse (singular) or infinite expansion behaviors for $\tau\rightarrow\pm\infty$ in every phase-space variable.

Now we will study the way this behavior gets modified (if at all) at the level of the noncommutative coherent state action (\ref{NCaction}).

\section{Quantum Noncommutative Bianchi I model}

Before promoting the classical pair $(a_i,\pi^i)$ directly to quantum operators we first make some pertinent considerations on the significance of noncommutativity in a cosmological context.

Because the scale factors are arbitrary and only have meaning relative to their values at some other epoch, then we will be interested in their relation with respect to their values during the Planck epoch. Thus, from our introductory remarks we will assume that the corresponding field observables $\hat a_i$ will dynamically inherit, as Heisenberg operators (see \cite{Ros07} for details of such mechanism), the noncommutativity of space (\ref{nchw}), which in turn should have been relevant when the size of the Universe was comparable to a Planck length $\ell_p$. This can be achieved through the commutator
\be\label{CNC}
[\hat a_i,\hat a_j]:=i\tilde\theta_{ij},\quad\tilde\theta_{ij}=\lambda^2\theta_{ij}
\ee
where $\lambda$ is a continuous parameter of units $L^{-1}$ proportional to the inverse of the (mean) radius of the Universe at some epoch and $|\theta_{ij}|\simeq\ell_p^2$ as before. Therefore we see that during the Planck epoch $\tilde\theta_{ij}\approx1$, meanwhile in the present epoch $\tilde\theta_{ij}\approx10^{-122}$, which reproduces the notion that noncommutative effects at cosmological scales are no longer observed.

Now to complete our algebra of operators we may implement the ordinary commutator between the $\hat a_i$ and their conjugate momenta $\hat\pi^i$ as
\be\label{CHW}
[\hat a_i,\hat\pi^j]:=i\hbar\rho\lambda^2\delta_i^j,
\ee
where the parameter $\rho$ has units $(ML/T)^{-1}$, and so $\lambda$ and $\rho$ are to be related through $(\rho\lambda)^{-1}\geq\hbar$ in order to be physically consistent magnitudes. The previous commutator is dimensionally correct recalling that the classical $\pi^i$'s have units $L^{-1}$. In particular, for $(\rho\lambda)^{-1}=\hbar$
\be
[\hat a_i,\hat\pi^j]=i\lambda\delta_i^j,
\ee
which for large scales recovers the commutative classical behavior of $\hat a_i$ and $\hat\pi^i$ in a similar fashion to (\ref{CNC}).

Therefore during calculations we may regard the parameters $\lambda$ and $\rho$ simply as factors meant to fix units, and return to their interpretation during the various epochs later.

After quantization, the Hamiltonian constraint (\ref{classconst}) can be identified with the Hermitian operator:
\be\label{constr}
\hat C_{grav}={\hat\pi}^i{\hat a_i}^2{\hat\pi}^i-{1\over2}({\hat\pi}^i{\hat a}_i)({\hat a}_j{\hat\pi}^j),
\ee
however, it will not be regarded as the quantum constraint. The latter will be obtained, according to expression (\ref{normalorder}), through normal ordering using appropriate noncommutative ladder operators.

Similarly to what was done with noncommutative coherent states in \S 4, we may now define the dimensionless operators
\be\label{Cncladder}
\begin{split}
\hat\Gamma_i:={1\over\sqrt{2\hbar\lambda\rho}}\big(\hat a_i+{1\over2\hbar\lambda^2\rho}\tilde\theta_{ij}\hat \pi^j+{i\over\lambda}\hat \pi^i\big),\\
\hat\Gamma_i^\dag:={1\over\sqrt{2\hbar\lambda\rho}}\big(\hat a_i+{1\over2\hbar\lambda^2\rho}\tilde\theta_{ij}\hat \pi^j-{i\over\lambda}\hat \pi^i\big),
\end{split}
\ee
so that by using (\ref{CNC}) and (\ref{CHW}) it can be verified that they satisfy the commutators
\be\label{NCcre-anni}
[\hat\Gamma_i,\hat\Gamma_j^\dag]=\delta_{ij},\quad [\hat\Gamma_i,\hat\Gamma_j]=[\hat\Gamma_i^\dag,\hat\Gamma_j^\dag]=0,
\ee
with the inverse forms
\be\label{inv}
\begin{split}
\hat a_i&=\sqrt{\hbar\lambda\rho/2}\bigg[\hat\Gamma_i+\hat\Gamma_i^\dag+\sum_j{i\tilde\theta_{ij}\over2\hbar\lambda\rho}(\hat\Gamma_j-\hat\Gamma_j^\dag)\bigg],\\
\hat\pi^i&=-i\lambda\sqrt{\hbar\lambda\rho/2}[\hat\Gamma_i-\hat\Gamma_i^\dag].
\end{split}
\ee

After substituting (\ref{inv}) into (\ref{constr}) and making repeated use of (\ref{NCcre-anni}), along with the antisymmetry of $\tilde\theta_{ij}$, we can arrive at the normal ordered form of $\hat C_{grav}$ which will be promoted to act as the quantum constraint $\hat\Phi_{grav}$, {\it i.e.}
\be\label{Bconstraint}
\begin{split}
:\hat C_{grav}:=&\hbar^2\lambda^4\rho^2\bigg[{1\over8}\sum_{i,j}(\hat\Gamma_i^2\hat\Gamma_j^2-2\hat\Gamma_i^{\dagger^2}\hat\Gamma_j^2+\hat\Gamma_i^{\dagger^2}\hat\Gamma_j^{\dagger^2})(1-2\delta_{ij})+{1\over2}\sum_i \hat\Gamma_i^{\dagger}\hat\Gamma_i\\
&-{i\over4\hbar\lambda\rho}\sum_{i,j}\tilde\theta_{ij}(\hat\Gamma_i^3-\hat\Gamma_i^{\dagger^2}\hat\Gamma_i-\hat\Gamma_i^\dagger\hat\Gamma_i^2+\hat\Gamma_i^{\dagger^3}-\hat\Gamma_i-\hat
\Gamma_i^\dagger)(\hat\Gamma_j-\hat\Gamma_j^\dagger)\\
&+\bigg({1\over4\hbar\lambda\rho}\bigg)^2\sum_{i,j,k}\tilde\theta_{ij}\tilde\theta_{ik}(\hat\Gamma_i^2-2\hat\Gamma_i^\dagger\hat\Gamma_i+\hat\Gamma_i^{\dagger^2}-1)(\hat\Gamma_j\hat\Gamma_k-2\hat\Gamma_j^\dagger\hat \Gamma_k+\hat\Gamma_j^\dagger\hat\Gamma_k^\dagger-\delta_{jk})\bigg]\\
&=\hat\Phi_{grav}.
\end{split}
\ee

Now, comparing (\ref{NCcre-anni}) with (\ref{caalg}) and according to the construction of the coherent states (\ref{nccs}), the corresponding coherent states for the Bianchi I anisotropic noncommutative quantum model are obtained from the transitive action
\be\label{CosNCCS}
|\underline{\gamma}\rangle=\exp\bigg[\sum_{i=1}^3(\gamma_i\hat\Gamma_i^\dag-\gamma_i^*\hat \Gamma_i)\bigg]|0\rangle=\prod_{i=1}^3e^{(\gamma_i\hat \Gamma_i^\dag-\gamma_i^*\hat \Gamma_i)}|0\rangle,
\ee
characterized by the eigenvalue equation
\be
\hat\Gamma_i|\underline\gamma\rangle=\gamma_i|\underline\gamma\rangle,\quad\gamma_i\in\mathbb{C}.
\ee

Making use of these elements and general expression (\ref{pathintf}) we can immediately write the corresponding constrained reproducing kernel that will act as the propagator of the theory
\be\label{Cpathintf}
\begin{split}
\langle\underline{\gamma}''|\mathbb{\hat E}|\underline{\gamma}'\rangle=\int_{\underline{\gamma}'}^{\underline{\gamma}''}\mathcal{D}\mu(\underline{\gamma},\underline{\gamma}^*)\int d\mu(\sigma)e^{i\int dt[{i\over2}(\underline{\gamma}^*\cdot\underline{\dot\gamma}
-\underline{\gamma}\cdot\underline{\dot\gamma}^*)-\sigma\Phi_{grav}(\underline{\gamma},\underline{\gamma}^*)]},
\end{split}
\ee
where $\Phi_{grav}(\underline{\gamma},\underline{\gamma}^*)=\langle\underline{\gamma}|\hat\Phi_{grav}|\underline{\gamma}\rangle$ is directly computed from (\ref{Bconstraint})
\be\label{Bconstraintclass}
\begin{split}
\Phi_{grav}(\underline{\gamma},\underline{\gamma}^*)=&\hbar^2\lambda^4\rho^2\bigg[{1\over8}\sum_{i,j}(\gamma_i^2\gamma_j^2-2\gamma_i^{*^2}\gamma_j^2+\gamma_i^{*^2}\gamma_j^{*^2})(1-2\delta_{ij})+{1\over2}\sum_i \gamma_i^*\gamma_i\\
&-{i\over4\hbar\lambda\rho}\sum_{i,j}\tilde\theta_{ij}(\gamma_i^3-\gamma_i^{*^2}\gamma_i-\gamma_i^*\gamma_i^2+\gamma_i^{*^3}-\gamma_i-
\gamma_i^*)(\gamma_j-\gamma_j^*)\\
&+\bigg({1\over4\hbar\lambda\rho}\bigg)^2\sum_{i,j,k}\tilde\theta_{ij}\tilde\theta_{ik}\big[(\gamma_i^2-2\gamma_i^*\gamma_i+\gamma_i^{*^2})(\gamma_j\gamma_k-2\gamma_j^* \gamma_k+\gamma_j^*\gamma_k^*)\\
&\qquad\qquad-(\gamma_i^2-2\gamma_i^*\gamma_i+\gamma_i^{*^2})\delta_{jk}-(\gamma_j\gamma_k-2\gamma_j^* \gamma_k+\gamma_j^*\gamma_k^*)\big]\bigg].
\end{split}
\ee

To recover physical insight we recall from (\ref{exval}) and using (\ref{inv}) that a pair of phase-space coordinates $(r_i,p_i)$ is obtained from the expectations
\be\label{Cexval}
\begin{split}
r_i&:=\langle\underline{\gamma}|\hat a_i|\underline{\gamma}\rangle=\sqrt{\hbar\lambda\rho/2}\bigg[\gamma_i+\gamma_i^*+\sum_j{i\tilde\theta_{ij}\over2\hbar\lambda\rho}(\gamma_j-\gamma_j^*)\bigg],\\
p_i&:=\langle\underline{\gamma}|\hat\pi^i|\underline{\gamma}\rangle=-i\lambda\sqrt{\hbar\lambda\rho/2}[\gamma_i-\gamma_i^*],\\
\end{split}
\ee
and by inverting them we can express the holomorphic coordinates in terms of physical variables as
\be\label{invclass}
\begin{split}
\gamma_i:={1\over\sqrt{2\hbar\lambda\rho}}\big(r_i+{1\over2\hbar\lambda^2\rho}\sum_j\tilde\theta_{ij}p_j+{i\over\lambda}p_i\big),\\
\gamma_i^*:={1\over\sqrt{2\hbar\lambda\rho}}\big(r_i+{1\over2\hbar\lambda^2\rho}\sum_j\tilde\theta_{ij}p_j-{i\over\lambda} p_i\big).
\end{split}
\ee

Then, by substituting (\ref{invclass}) in (\ref{Bconstraintclass}) and after lengthy algebraic simplifications we obtain
\be\label{NCclassconst}
\begin{split}
\Phi_{grav}(\underline{r},\underline{p})=&\sum_ip_i^2r_i^2-{1\over2}\bigg(\sum_ip_ir_i\bigg)^2+{\hbar\lambda\rho\over4}\sum_i(p_i^2+\lambda^2r_i^2)+{\lambda\over4}\sum_{i,j}{\tilde\theta_{ij}}p_ir_j\\
&-{1\over16\hbar\lambda\rho}\sum_{i,j,k}\tilde\theta_{ij}\tilde\theta_{ik}p_jp_k(1-2\delta_{jk}),
\end{split}
\ee
where the first two terms are identified as the classical Hamiltonian constraint (\ref{classconst}), whereas the other terms correspond to quantum and noncommutative corrections.

An important element to note here is the presence of a purely noncommutative correction linear in $\tilde\theta_{ij}$, in the form of an angular momentum. According to the results in \cite{Min09} we can already expect this term to be relevant in the stability of the solutions and central in preventing the collapse in the deep Planckian sector.

%To further simplify calculations we will study the case where noncommutativity is present only between $\hat a_1$ and $\hat a_2$, \emph{i.e}, $\tilde\theta_{12}=-\tilde\theta_{21}=\tilde\theta\neq0$, in which case

\section{Path Integral action, dynamics and $\mu$-parameters}

Similarly to what was done to arrive at expression (\ref{NCaction}), the action in the path integral (\ref{Cpathintf}) can be rewritten in terms of physical variables using (\ref{invclass})
\be\label{CNCaction}
S={1\over\lambda^2\rho}\int d\tau\big[\sum_{i}\big(p_i{\dot r_i}+\sum_j{\tilde\theta_{ij}\over2\hbar\lambda^2\rho}p_i{\dot p_j}\big)-\lambda^2\rho\hbar\sigma\Phi_{grav}({\underline q},{\underline p})\big],
\ee
and noticing that in order to recover the classical equations of motion (\ref{classdyn}), controlled only by the first two terms in (\ref{NCclassconst}), \emph{i.e.} when $\hbar=\tilde\theta=0$, we need to identify $\sigma={1\over\hbar\lambda^2\rho}$. Observe that this involves integrating over the measure $d\mu({1\over\hbar\lambda^2\rho})$ in the transition amplitude (\ref{Cpathintf}) which makes the final result independent of scales.

As is well known in the theory of Feynman integrals, the trajectories for which $\delta S=0$ are the ones that most contribute to the propagator, such trajectories are solutions to the Euler-Lagrange equations of (\ref{CNCaction}) which, by using (\ref{NCclassconst}), are immediately seen to be
\be\label{dynamics1}
\begin{split}
\dot r_i&=r_i(2p_ir_i-\sum_jp_jr_j)+{\hbar\lambda\rho\over2}p_i+{\lambda\over4}\sum_j\tilde\theta_{ij}r_j+{1\over8\hbar\lambda\rho}\sum_{j,k}\tilde\theta_{ij}\tilde\theta_{jk}p_j(1-2\delta_{ik})-{1\over\hbar\lambda^2\rho}\sum_j\tilde\theta_{ij}\dot p_j,\\
\dot p_i&=-p_i(2p_ir_i-\sum_jp_jr_j)-{\hbar\lambda^3\rho\over2}r_i+{\lambda\over4}\sum_j\tilde\theta_{ij}p_j.
\end{split}
\ee

Instead of working with the previous equations it will prove more convenient to work with dimensionless expressions. This can be done first by substituting $\tilde\theta_{ij}=\lambda^2\theta_{ij}$, then removing the units of any $p_i$ and any $\tau$ derivative through the change of variables $p_i=\lambda\tilde p_i$ and $\dot f=\lambda f'$, where $f$ is arbitrary. After carrying this out one arrives at the equivalent dimensionless system
\be\label{dynamics2}
\begin{split}  r_i'&=r_i(2\tilde p_ir_i-\sum_j\tilde p_jr_j)+{\hbar\lambda\rho\over2}\tilde p_i+{\lambda^2\over4}\sum_j\theta_{ij}r_j+{\lambda^3\over8\hbar\rho}\sum_{j,k}\theta_{ij}\theta_{jk}\tilde p_j(1-2\delta_{ik})-{\lambda\over\hbar\rho}\sum_j\theta_{ij}\tilde p_j',\\
\tilde p_i'&=-\tilde p_i(2\tilde p_ir_i-\sum_j\tilde p_jr_j)-{\hbar\lambda\rho\over2}r_i+{\lambda^2\over4}\sum_j\theta_{ij}\tilde p_j.
\end{split}
\ee
with this normalization of units it is possible to see that the limit $\lambda\rightarrow0$ properly reproduces the classical equations of motion, whereas this limit was ill defined in equations (\ref{dynamics1}).

By defining the dimensionless "$\mu$-parameters"
\be
\mu_{\hbar}(\lambda):=\hbar\lambda\rho\leq1,\quad \mu_{ij}^\theta(\lambda):=\lambda^2\theta_{ij}\leq1,
\ee
which characterize the quantum and planckian scales respectively, the differential system (\ref{dynamics2}) takes the form
\be\label{dynamics3}
\begin{split}  r_i'&=r_i(2\tilde p_ir_i-\sum_j\tilde p_jr_j)+{\mu_\hbar\over2}\tilde p_i+{3\over4}\sum_j\mu_{ij}^\theta r_j+\sum_j{\mu_{ij}^\theta\over \mu_\hbar}\tilde p_j\bigg(2\tilde p_jr_j-\sum_k\tilde p_kr_k\bigg)+{1\over8}\sum_{j,k}{\mu_{ij}^\theta \mu_{jk}^\theta,\over \mu_\hbar}\tilde p_j(1+2\delta_{ik}),\\
\tilde p_i'&=-\tilde p_i(2\tilde p_ir_i-\sum_j\tilde p_jr_j)-{\mu_\hbar\over2}r_i+{1\over4}\sum_j\mu_{ij}^\theta\tilde p_j.
\end{split}
\ee
and the constraint (\ref{NCclassconst}) now reads
\be\label{NCclassconst2}
\begin{split}
\tilde\Phi_{grav}(\underline{r},\underline{\tilde p})=&\sum_i\tilde p_i^2r_i^2-{1\over2}\bigg(\sum_i\tilde p_ir_i\bigg)^2+{\mu_\hbar\over4}\sum_i(\tilde p_i^2+\lambda^2r_i^2)+{1\over4}\sum_{i,j}\mu_{ij}^\theta\tilde p_ir_j-{1\over16}\sum_{i,j,k}{\mu_{ij}^\theta \mu_{ik}^\theta,\over \mu_\hbar}\tilde p_j\tilde p_k(1-2\delta_{jk}),
\end{split}
\ee
Observe that the pure quantum $\mu_\hbar$ and noncommutative $ \mu_{ij}^\theta$ sectors can now be clearly distinguished in (\ref{dynamics3}) and (\ref{NCclassconst2}), as well as the interplay between them trough the quotient sectors ${\mu_{ij}^\theta\over \mu_\hbar},{\mu_{ij}^\theta \mu_{jk}^\theta\over \mu_\hbar}$, which in turn correspond to newly derived $\mu$-parameters.

Because of their $\lambda$-scale dependence, a strict hierarchy can be established among the different $\mu$-parameters for all
epochs, making them analogous to the running couplings of field theories. Thus the renormalization group for this
quantum noncommutative Bianchi I cosmology is characterized by the couplings
$\mu_\hbar,\mu_{ij}^\theta,{\mu_{ij}^\theta\over \mu_\hbar},{\mu_{ij}^\theta \mu_{jk}^\theta\over \mu_\hbar}$. Within this context the fact
that this group has an infrared stable point, \emph{i.e.} when $\lambda\rightarrow0$, confirms that the classical Bianchi I
cosmology results as an effective theory of our model.

\section{Bounce-like Paths}

As the classical symmetries are broken in the planckian regime, and are recovered only in the limit $\lambda\rightarrow0$,
the quantities $\tilde p_ir_i$ are no longer constants of motion and analytical solutions for this highly coupled nonlinear
differential system are not easy to come by. In this section we will show that a specific type of nonsingular solutions, namely a family of bounce-like trajectories for the volume scaling function $v:=r_1r_2r_3$, are allowed by the equations of motion (\ref{dynamics3}). To that purpose we give the following:

\begin{Def}[Bounce]
The differential system (\ref{dynamics3}) will be said to have a bounce inside an interval $B=(\tau_B-\epsilon,\tau_B+\epsilon)$ when all the momenta undergo a sign change for values $\tau\in $B, where in the simplest case this occurs simultaneously meaning that at $\tau=\tau_B$ the momenta satisfy $\tilde p_i(\tau_B)=0$.
\end{Def}

The previous definition is not sufficient, however, to guarantee all the properties we look for in the solutions, where for the simplest case the volume scaling function $v(\tau)=r_1(\tau)r_2(\tau)r_3(\tau)$ together with its first and second derivatives should also satisfy
\be\label{conds}
\begin{split}
&v(\tau_B)>0,\\
&v'(\tau_B)=0,\\
&v''(\tau_B)>0.
\end{split}
\ee
Therefore by using (\ref{dynamics3}) to explicitly calculate the above derivatives, where for simplicity sake we fix $\mu_{12}^\theta=\mu_{23}^\theta=\mu_{31}^\theta=\mu_\theta$, and substituting the bounce definition, it is not very difficult to show that at $\tau=\tau_B$ the last two expressions in (\ref{conds}) yield
\be\label{conds2}
v'(\tau_B)={3\mu_\theta\over4}\sum_{(ijk)}r_i^2\big(r_j-r_k\big)=0,
\ee
where $(ijk)$ are the cyclic permutations of indices $(123)$, and
\be\label{conds3}
v''(\tau_B)=\bigg[{\mu_\hbar\over2}\sum_ir_i^2-{\mu_\theta^2\over8}\sum_{(ijk)}{r_i\over r_jr_k}\big(9r_i-14(r_j+r_k)\big)-{3\mu_\hbar^2\over4}-{63\mu_\theta^2\over8}\bigg]v(\tau_B)>0.
\ee
We then observe from (\ref{conds2}) that the simplest solution corresponds to a bounce for $r_1=r_2=r_3=r_B$ that, when substituted in (\ref{conds3}), leads to the greatly simplified lower bound
\be\label{conds4}
{3\mu_\hbar\over 2}r_B^2>{3\over4}(\mu_\hbar^2+\mu_\theta^2).
\ee

To better understand the implications of these results observe first that the condition to extremize the volume is of pure noncommutative nature, in fact, it originates from the noncommutative term of angular momentum in (\ref{NCclassconst}). This confirms our remark in the last paragraph of \S7, thus establishing the relevance of the angular momentum in controlling the bounce and providing more evidence that noncommutative effects are manifest dynamically. On the other hand, condition (\ref{conds4}) expresses the infimum for $r_B$ at all scales above which a bounce occurs. But more interesting is the fact that in the Planckian regime when $\mu_\hbar=\mu_\theta=1$ this condition reduces simply to $r_B>1$, or equivalently that the bounce has to occur for scales greater than one cubic Planck lenght.

For a deeper analysis of the behavior of the system (\ref{dynamics3}) around $\tau_B$ and to determine if \emph{bounce}-like
solutions are admissible, we apply the  method of matched asymptotic expansions often used in
%critical layers in shear flows
boundary-layer problems \cite{Nayfeh,Kev96}. Contrary to the usual procedure, we are not interested in finding the first terms of
an asymptotic analytical solution for (\ref{dynamics3}), but to show that the zeroth order of the expansion near the Planckian regime
is compatible with a bounce-like behavior. To do this we start in a scale one order of magnitude larger than the Planckian {\it i.e} $\lambda\approx10^{32}cm^{-1}$, or $\mu_\hbar=10^{-1}=\varepsilon$ and $\mu_\theta=10^{-2}=\varepsilon^2$, in a manner such that the quantum and noncommutative corrections can be handled perturbatively around the well known classical solutions, but where noncommutative effects from the Planckian regime remain relevant for the evolution.
Thus the system (\ref{dynamics3}) becomes:
\be\label{dynamics4}
\begin{split} r_i'&=r_i(\tilde p_ir_i-\tilde p_jr_j-\tilde p_kr_k)+{\varepsilon}\tilde p_j(\tilde p_jr_j-\tilde p_kr_k-\tilde p_ir_i)-{\varepsilon}\tilde p_k(\tilde p_kr_k-\tilde p_ir_i-\tilde p_jr_j)\\
     &\quad+{\varepsilon\over2}\tilde p_i+{3\varepsilon^2\over4}r_j -{3\varepsilon^2\over4}r_k+ \mathcal{O}(\varepsilon^3),\\
\tilde p_i'&=-\tilde p_i(\tilde p_ir_i-\tilde p_jr_j-\tilde p_kr_k)-{\varepsilon\over2}r_i+{\varepsilon^2\over4}\tilde p_j
             -{\varepsilon^2\over4}\tilde p_k + \mathcal{O}(\varepsilon^3),
\end{split}
\ee
where the indices $(i,j,k)$ are cyclically ordered.
%It is evident that this choice of scales still allows us to recover the \emph{classical} behavior
%in the limit $\varepsilon\rightarrow0$.\\

\begin{figure}[h!]
\centering
\includegraphics[width=7cm]{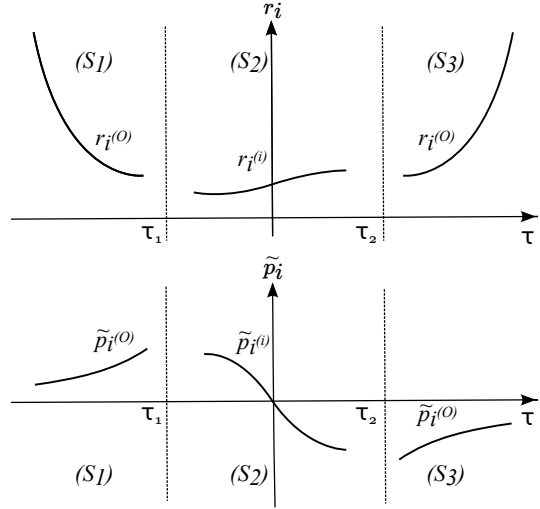}
\caption{\footnotesize{The expected behavior of the bounce-like solutions is shown. The evolution of $r_i$ (top) and $\tilde{p}_i$ (bottom)
is separated in the three regions of interest, detailing their general asymptotic properties.}}\label{regs}
\end{figure}

The type of solutions for (\ref{dynamics4}) that are our main concern are such that their behavior away from Planckian scales corresponds to that of a classical collapse in the past which transits to a classical expansion in the future. From the way in which the linear terms appear in the system, an oscillatory behavior is expected in the region where such terms are dominant, namely for small values of $r_i$ and $p_i$. Imposing the bounce criteria that we introduced at the beginning of this section requires the solutions in the transition region, when going from collapse to expansion, to allow for the change of sign in the $p_i$ while keeping the $r_i$ positive, as pictured in Fig(\ref{regs}). \\

Thus as a first step we divide the dynamical phase space $(r,\tilde{p},\tau)$ in three regions: (I) $S_1=\{(r,\tilde{p},\tau)|\tau<\tau_1\}$, (II)
$S_2=\{(r,\tilde{p},\tau)|\tau_1<\tau<\tau_2\}$ and (III) $S_3=\{(r,\tilde{p},\tau)|\tau>\tau_2\}$, where we assume that outer (\emph{classical}) solutions
will be in regions $S_1$ and $S_3$ and an inner solution for region $S_2$. To simplify the calculations we will specialize
to the symmetrical case around the origin: $\tau_2=-\tau_1=\tilde{\tau}$.\\

For each region we have the asymptotic forms:
\be \label{asymp}
 \begin{split}
  r_j^{(A)}(t)=  \mathsf{R}_{j}^{(A)} + \mathcal{O}(\varepsilon),\\
  \tilde{p}_{j}^{(A)}(t)= \tilde{\mathsf{P}}_{j}^{(A)} + \mathcal{O}(\varepsilon),
 \end{split}
\ee
%\be \label{asymp}
% \begin{split}
%  r_j^{(A)}(t)= \mathsf{r}_{j}^{(A)} + \varepsilon \mathsf{R}_{j}^{(A)} + \mathcal{O}(\varepsilon^2),\\
%  \tilde{p}_{j}^{(A)}(t)= \tilde{\mathsf{p}}_{j}^{(A)} + \varepsilon \tilde{\mathsf{P}}_{j}^{(A)} + \mathcal{O}(\varepsilon^2),
% \end{split}
%\ee
where the upperscrit $(A)$ denotes whether the solution is outer ($O$) or the inner ($i$).

\subsection*{Outer solution}
For regions $S_1$ and $S_3$, i.e. outside the planckian region, the solutions are expected to behave as the \emph{classical} ones.
Then to zeroth order the equations (\ref{dynamics4}) read:
%\be\label{outsol}
%\begin{split} \mathsf{r}_1'&=\mathsf{r}_1(\tilde{\mathsf{p}}_1\mathsf{r}_1-\tilde{\mathsf{p}}_2\mathsf{r}_2-\tilde{\mathsf{p}}_3\mathsf{r}_3),\\
%\mathsf{r}_2'&=\mathsf{r}_2(\tilde{\mathsf{p}}_2\mathsf{r}_2-\tilde{\mathsf{p}}_1\mathsf{r}_1-\tilde{\mathsf{p}}_3\mathsf{r}_3),\\
%\mathsf{r}_3'&=\mathsf{r}_3(\tilde{\mathsf{p}}_3\mathsf{r}_3-\tilde{\mathsf{p}}_1\mathsf{r}_1-\tilde{\mathsf{p}}_2\mathsf{r}_2),\\
%\tilde{\mathsf{p}}_1'&=-\tilde{\mathsf{p}}_1(\tilde{\mathsf{p}}_1\mathsf{r}_1-\tilde{\mathsf{p}}_2\mathsf{r}_2-\tilde{\mathsf{p}}_3\mathsf{r}_3),\\
%\tilde{\mathsf{p}}_2'&=-\tilde{\mathsf{p}}_2(\tilde{\mathsf{p}}_2\mathsf{r}_2-\tilde{\mathsf{p}}_1\mathsf{r}_1-\tilde{\mathsf{p}}_3\mathsf{r}_3),\\
%\tilde{\mathsf{p}}_3'&=-\tilde{\mathsf{p}}_3(\tilde{\mathsf{p}}_3\mathsf{r}_3-\tilde{\mathsf{p}}_1\mathsf{r}_1-\tilde{\mathsf{p}}_2\mathsf{r}_2),
%\end{split}
%\ee
\be\label{outsol}
\begin{split} \mathsf{R}_i'&=\mathsf{R}_i(\tilde{\mathsf{P}}_i\mathsf{R}_i-\tilde{\mathsf{P}}_j\mathsf{R}_j-\tilde{\mathsf{P}}_k\mathsf{R}_k),\\
\tilde{\mathsf{P}}_i'&=-\tilde{\mathsf{P}}_i(\tilde{\mathsf{P}}_i\mathsf{R}_i-\tilde{\mathsf{P}}_j\mathsf{R}_j-\tilde{\mathsf{P}}_k\mathsf{R}_k),
\end{split}
\ee
which have the usual classical solutions:
\be \label{outsol}
\begin{split}
   \mathsf{R}_i(\tau)&=\mathsf{R}_i(\tau_0)e^{(\chi_i-\chi_j-\chi_k)(\tau-\tau_0)}, \\
   \tilde{\mathsf{P}}_i(\tau)&=\tilde{\mathsf{P}}_i(\tau_0)e^{-(\chi_i-\chi_j-\chi_k)(\tau-\tau_0)}
\end{split}
\ee
where as before the indices $(i,j,k)$ are cyclic and $\chi_i:= \mathsf{R}_i(\tau_0)\tilde{\mathsf{P}}_i(\tau_0)$ with $\tau_0$ corresponding to a time in the remote past for $S_1$ and in the distant future for $S_3$. Where a \emph{classical} collapse (expansion)
for region $S_1$ ($S_3$) occurs for: $\mathsf{R}_i(\tau_1)>0$ ($\mathsf{R}_i(\tau_2)>0$) and $\tilde{\mathsf{P}}_i(\tau_1)>0$
($\tilde{\mathsf{P}}_i(\tau_2)<0$). Which implies that a necessary condition for a bounce-like solution of the volume scaling function is for the momenta to change sign inside
$S_2$.

\subsection*{Inner solution}
As the noncommutative terms $\mathcal{O}(\varepsilon^2)$ in (\ref{dynamics4}) are expected to become important inside $S_2$, it is reasonable to scale down the variables to that same order to assert the relevant dynamics in this region, \emph{i.e.} $(r_i,\tilde{p}_i) \rightarrow (\varepsilon^2 \mathsf{r}_i, \varepsilon^2 \mathsf{p}_i)$. To further simplify the equations resulting from this scaling we also introduce the time variable
$t=\varepsilon(\tau-\tau_B)$, thus the system (\ref{dynamics4}) in the inner region reduces to:
\be\label{dynamics5}
\begin{split} \frac{d\mathsf{r}_i}{dt}&={1\over2}\mathsf{p}_i+{3\varepsilon\over4}\mathsf{r}_j
             -{3\varepsilon\over4}\mathsf{r}_k + \mathcal{O}(\varepsilon^3),\\
\frac{d\mathsf{p}_i}{dt}&=-{1\over2}\mathsf{r}_i+{\varepsilon\over4}\mathsf{p}_j
             -{\varepsilon\over4}\mathsf{p}_k + \mathcal{O}(\varepsilon^3),
\end{split}
\ee
where the indices $(i,j,k)$ are again cyclically ordered.
The solutions to this linear system can be obtained in closed forms, however, we only present the first terms of their asymptotic expansion. The six orthogonal eigenvectors at leading order
$\varepsilon$ are:
\be\label{solutions1}
\begin{split}
\mathsf{w}_1(t)=&(\cos(\nu_0 t),\cos(\nu_0 t),\cos(\nu_0 t),
                -\sin(\nu_0t),-\sin(\nu_0t),-\sin(\nu_0t)),\\
\mathsf{w}_2(t)=&(\sin(\nu_0t),\sin(\nu_0t),\sin(\nu_0t),\cos(\nu_0t),
               \cos(\nu_0t),\cos(\nu_0t)), \\
\mathsf{w}_3(t)=&(\cos(\nu_{+}t)+{\sqrt 3}\sin(\nu_+t),
              \cos(\nu_{+}t)-{\sqrt 3}\sin(\nu_+t),-2\cos(\nu_{+}t),\\
              &-{\sqrt 3}\xi_-\cos(\nu_{+}t)-\xi_-\sin(\nu_+t),{\sqrt 3}\xi_-\cos(\nu_{+}t)- \xi_-\sin(\nu_+t),2\xi_-\sin(\nu_{+}t)),\\
\mathsf{w}_4(t)=&(-\sin(\nu_{+}t)+{\sqrt 3}\cos(\nu_+t),
              -\sin(\nu_{+}t)-{\sqrt 3}\cos(\nu_+t), 2\sin(\nu_{+}t),\\
              &{\sqrt 3}\xi_-\sin(\nu_{+}t)-\xi_-\cos(\nu_+t),-{\sqrt 3}\xi_-\sin(\nu_{+}t)- \xi_-\cos(\nu_+t),2\xi_-\cos(\nu_{+}t)), \\
\mathsf{w}_5(t)=&(\cos(\nu_{-}t)-{\sqrt 3}\sin(\nu_-t),
              \cos(\nu_{-}t)+{\sqrt 3}\sin(\nu_-t),-2\cos(\nu_{-}t), \\
              &{\sqrt 3}\xi_+\cos(\nu_{-}t)-\xi_+\sin(\nu_-t),-{\sqrt 3}\xi_+\cos(\nu_{-}t)- \xi_+\sin(\nu_-t),2\xi_+\sin(\nu_{-}t)),\\
\mathsf{w}_6(t)=&(-\sin(\nu_{-}t)-{\sqrt 3}\cos(\nu_-t),
              -\sin(\nu_{-}t)- {\sqrt 3}\cos(\nu_-t),2\sin(\nu_{-}t), \\
              &-{\sqrt 3}\xi_+\sin(\nu_{-}t)-\xi_+\cos(\nu_-t),{\sqrt 3}\xi_+\sin(\nu_{-}t)-\xi_+\cos(\nu_-t),2\xi_+\cos(\nu_{-}t)),
\end{split}
\ee
where  $\xi_{\pm}=\frac{4\pm\sqrt{3}\varepsilon}{4\mp\sqrt{3}\varepsilon}$,
$\nu_\pm=\frac{1}{2}\pm\frac{\sqrt{3}}{2}\varepsilon$ and $\nu_0=\frac{1}{2}$.

A solution that satisfies the prescription that the $\mathsf{p}_i$ change sign around the origin with $\mathsf{r}_i$ positive can be obtained with the
linear combination:
%From these independient solutions we can indeed construct
%an inner solution where the momenta $\tilde{p}_i$ do change its sign around the origin and do it so we also impose an initial sign for
%the $q_i$. Namely, if we choose a $q_i>0$, we have the inner solution:
\be\label{solutions2}
\begin{split}
\big(\underline{\mathsf{r}},\underline{\mathsf{p}}\big)_{S_2}^{(i)} =&
          \mathsf{A} \mathsf{w}_1 +  \mathsf{B} \Big( \xi_+\mathsf{w}_3  +\xi_-\mathsf{w}_5\Big)
         + \mathsf{C} \Big( \xi_+\mathsf{w}_4- \xi_-\mathsf{w}_6\Big)
\end{split}
\ee
where $(\underline{\mathsf{r}},\underline{\mathsf{p}})^{(i)}=(\mathsf{r}_1^{(i)},\mathsf{r}_2^{(i)},\mathsf{r}_3^{(i)},\mathsf{p}_1^{(i)},\mathsf{p}_2^{(i)},\mathsf{p}_3^{(i)})$ and
\be
\begin{split}
\mathsf{A}&=\frac{1}{3}\big(\mathsf{r}_1^{(i)}(0)+\mathsf{r}_2^{(i)}(0)+\mathsf{r}_3^{(i)}(0)\big),\\
\mathsf{B}&=\frac{1}{6}\big(\mathsf{r}_1^{(i)}(0)+\mathsf{r}_2^{(i)}(0)-2\mathsf{r}_3^{(i)}(0)\big),\\
\mathsf{C}&=\frac{1}{2\sqrt{3}}\big(\mathsf{r}_1^{(i)}(0)-\mathsf{r}_2^{(i)}(0)\big).
\end{split}
\ee

It is important to stress that the range of validity of this solution is confined to values $-\varepsilon<t<\varepsilon$, but that its qualitative properties persist throughout all the three regions on to the exponential regime.

From solution (\ref{solutions2}) we have various possible scenarios depending on the values of $r_i(0)$. In the isotropic case it simplifies only to the first expression in (\ref{solutions1}) meaning a unique evolution of the scale factors and also their conjugate momenta. In the more general scenario of different values for the three $r_i(0)$ the oscillation will be a linear combination of periodic functions with three different frequencies.
Also note that for the limit $\varepsilon\rightarrow0$ in (\ref{dynamics5}), which corresponds to neglect all the noncommutative terms, the system decouples into three identical oscillators causing the scale factors to evolve independently. The solution of this system can very well describe the behavior of the isotropic case near the bounce as the frequency is the same, but not in the arbitrary anisotropic case.
This method could now be extended to match both inner and outer solutions, through an intermediate solution, to obtain an approximate analytical solution valid for all $\tau$, however this goes beyond the scope of this work. To show the validity of our considerations and
solutions (\ref{outsol}) and (\ref{solutions2}), we recur to numerical solutions.

%We are not interested in finding the exact $\tilde t$ where we will do the matching of the inner and outer solutions. To do that usually
%requires to compute at least the next term of the inner and outer solutions. But there are some points that must be considered.
%
%Classical admisable inner and outer solutions can be matched as long all of them have positive values for $q_i$ and share the relative
%sign of $\tilde q_i$ on each side. But it is possible to construct non-physical admisable outer solutions with $q_i<0$ that can be matched
%with an inner solution if we let the matching occur for $\tau_2=\varepsilon t_2 > \frac{\pi}{2}$. This can kind of unphysical behavior was
%observed for numerical simulations with some critical initial values.

\section{Numerical Results}

\begin{figure}[h!]
  \includegraphics[width=10cm]{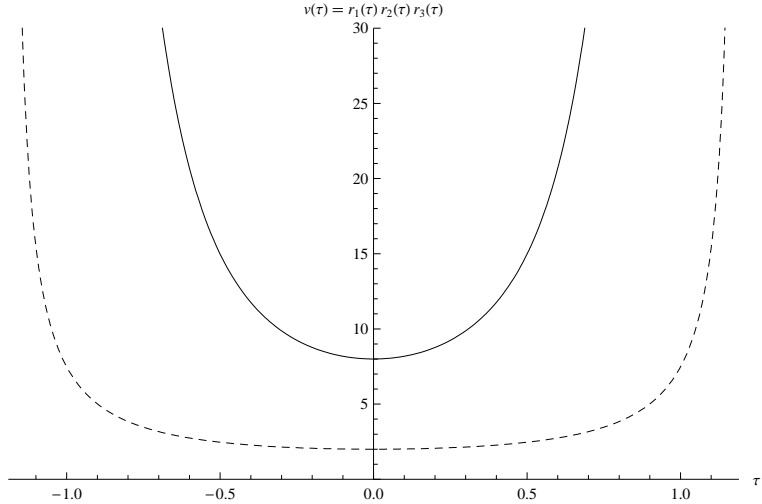}\\
  \caption{\footnotesize{Plot of the volume scaling function (solid) and scale factors (dashed) in the symmetric case $r_B=2$. The collapsing solution coming from the left bounces at the origin to evolve along an expanding trajectory to the right. The evolution of the scale factors is the same for the three cases.}}\label{f1}
\end{figure}

\begin{figure}[h!]
  \includegraphics[width=10cm]{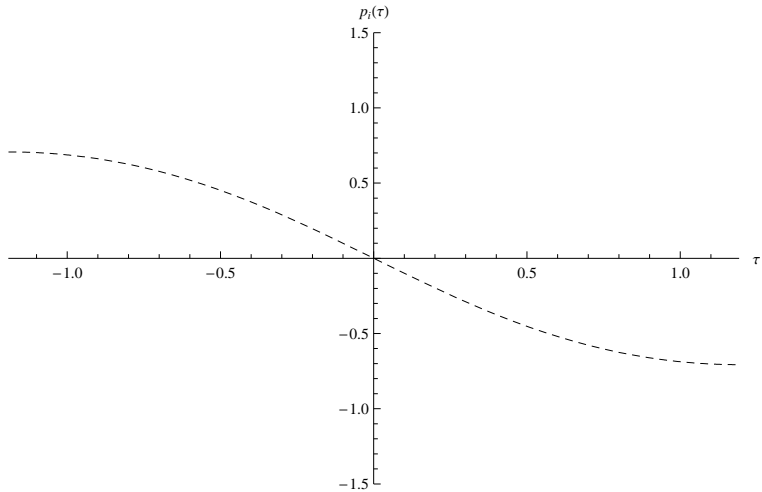}\\
  \caption{\footnotesize{Plot of the momenta conjugate to the $r_i$ from Fig.(\ref{f1}). The three momenta follow the same solution with the bounce occuring when they intersect the origin.}}\label{f2}
\end{figure}

In what follows all our simulations were obtained for values of $\mu_\hbar=\mu_{12}^\theta=\mu_{23}^\theta=\mu_{31}^\theta=1$ which corresponds to the deep Planckian regime. As a first check for our previous assumptions we provide the plot associated to the most symmetric case, when $r_1=r_2=r_3=r_B$ at the bounce point. Fig.(\ref{f1}) corresponds to the numerical solution of the volume scale function and the scale factors governed by the equations of motion (\ref{dynamics3}) when $r_B>1$. In this case the expectation of the three scale factors follows the same trajectory and are undistinguishable, which coincides with our analysis of the inner solution (\ref{solutions2}). All the scale factors evolve equally around the bounce before the exponential regime rapidly takes over with a similar situation occurring for the momenta in Fig.(\ref{f2}). Therefore the numerical solution confirms the simplest type of bounce hypothesized in the previous section.

\begin{figure}[h]
  \includegraphics[width=10cm]{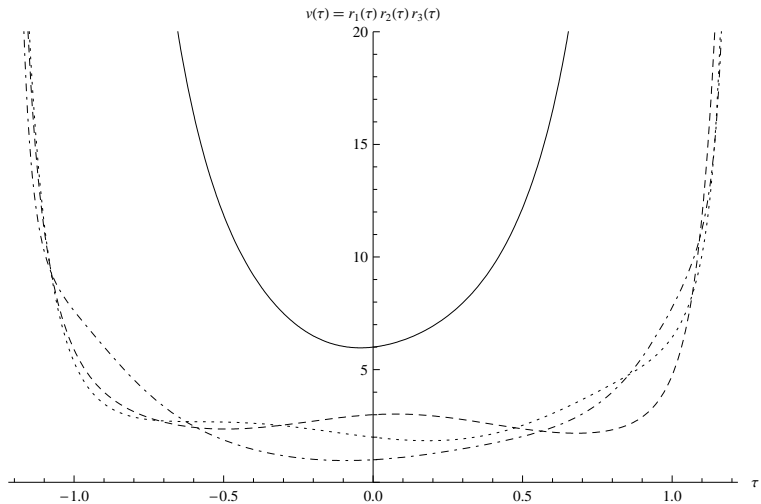}\\
  \caption{\footnotesize{Plot of the volume scaling function (solid) and scale factors $r_1$ (dashed), $r_2$ (dotted), $r_3$ (dotdashed) in the anisotropic case with values at the bounce $r_1=3, r_2=2,r_3=1$. The heterogenous oscillatory regime generated by the noncommutativity drives the evolution of the scale factors to coalesce into one asymptotic trajectory.}}\label{f3}
\end{figure}

A more interesting situation can be obtained by allowing anisotropy among the scale factors as exemplified in Fig.(\ref{f3}). While the bounce of the volume is clearly present and not too different from the previous case, a richer scenario occurs with the independent scale factors. Here, noncommutativity induces a secondary effect in the evolution by producing irregular oscillations of the trajectories around the bounce in the way predicted by the inner solutions obtained in (\ref{solutions2}), where contributions from frequencies $\nu_\pm$ are now also present. Note that these noncommutative oscillations are such that the solutions for the three scale factors tend towards a single trajectory in both branches until a point where, if the solution is continued in the vertical axis, no remnant of the original anisotropy is noticeable. This is a very appealing feature in the sense that it provides with a plausible mechanism to produce isotropic cosmologies (at large scales) in the far future out of anisotropic conditions in the remote past.

\begin{figure}[h]
  \includegraphics[width=10cm]{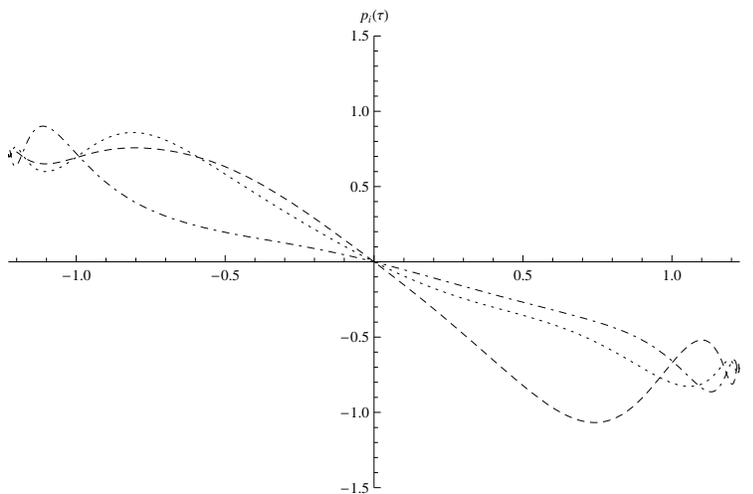}\\
  \caption{\footnotesize{Plot of the momenta conjugate to the $r_i$ of Fig.(\ref{f3}). The amplitude of the oscillations is suppressed as the system evolves away from the bounce.}}\label{f4}
\end{figure}

\begin{figure}[h]
  \includegraphics[width=10cm]{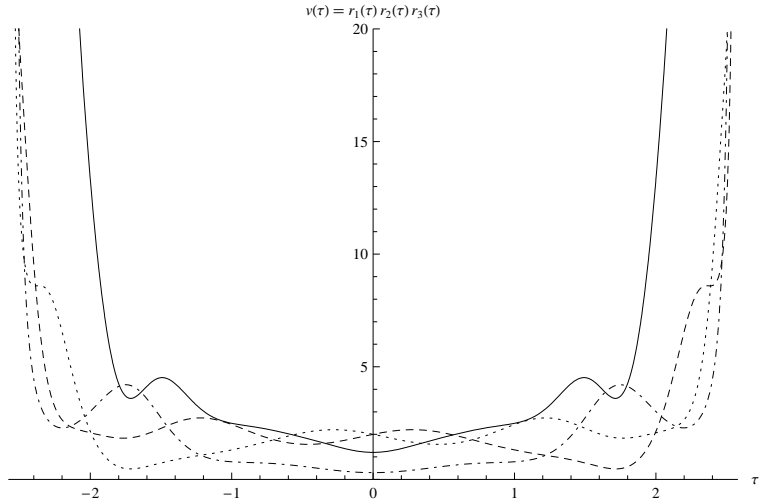}\\
  \caption{\footnotesize{Plot of the volume scaling function (solid) and scale factors $r_1$ (dashed), $r_2$ (dotted), $r_3$ (dotdashed) in the anisotropic case with values at the bounce $r_1=2, r_2=2,r_3=0.3$. More than one minima of the volume can be observed with only the central minimum satisfying the condition $p_i=0$.}}\label{f5}
\end{figure}

The third numerical example shows that bounce-like trajectories can display more than one minimum. In Fig.(\ref{f5}) we present the bounce-like solution obtained for anisotropic values of $r_i$ at the bounce such that the volume scale function remains above one Planck volume, yet the expectation of one scale factor can take sub-Planckian values. The evolution of the volume scaling function presents various minima, however, only the central minimum fulfills the bounce criteria, \emph{i.e.} $p_i=0$. Similar to the previous case, the three scale factors are seen to evolve towards an average trajectory.

\section{Discussion and Conclusions}

A set of coherent states were built for a noncommutative quantum Bianchi I anisotropic cosmology, which in turn were used to circumvent the absence of a simultaneous set of configuration observables. Instead, the chosen dynamical variables were the expectation values of the scale factors and momenta operators in the coherent base. By extending known methods of path integrals for constraint systems in terms of coherent states, and taking advantage of the property that noncommutative expressions have the same functional form as their commutative counterparts when written in holomorphic variables \cite{Ju10}, we obtained a form for the propagator of the theory and, with it, the corresponding modified noncommutative action.

As the dynamical variables ($r_i,p_i$) represent the center of gaussian states, a steepest descent analysis of the action was performed to obtain the approximate description of the full quantum states evolution, enhanced by the fact that the covariant Husimi Q-symbol for the path integral is more fit for our semiclassical scenario, \emph{cf.} \cite{Bar01}. Thus we obtained the corresponding equations of motion from the effective Hamiltonian in the path integral action where, by keeping all the units of the theory and introducing the dimensionless $\mu$-parameters, we were able to clearly differentiate the macroscopic dominant contributions from the pure quantum and noncommutative sectors as well as their mixed sectors. Here we must emphasize an important feature of our Hamiltonian constraint, that is confirmed simply by power counting, which is that the noncommutative corrections coming from our selection of dynamical variables are not originated in a Bopp map of the classical constraint, as is usually the case in other noncommutativity related literature.

From the perspective of an effective theory, our final expressions for the Hamiltonian (\ref{NCclassconst2}) and the equations of motion (\ref{dynamics3}) were shown to flow towards a stable infrared point, which indeed corresponds to the classical model. By changing the value of the $\lambda$-scale in the couplings of the theory, one can analyze the relevance of the noncommutative and quantum effects during different epochs. It is interesting to note that, although, the purely noncommutative effects die off relatively fast for scales larger than a few Planck lengths (equivalently, for $\lambda<1$), the mixed sector ${\mu_{ij}^\theta/\mu_\hbar}$ induces a noncommutative-quantum mechanical interaction that persists through such scales. This is a novelty of our model which gives hope to probe noncommutative effects at intermediate scales between the quantum regime and the Planckian one, a gap known in particle physics as the great desert for its attributed absence of new physics.

By providing a physically admissible definition for the bounce of the system we showed that it was compatible with the existence of a minimum for the volume scaling function. The particular scenario analyzed in detail represented an evolution where the volume collapses for cosmological times $\tau<\tau_B$, reaches its minimum value at $\tau_B$ and then expands for $\tau>\tau_B$. In order to satisfy the supplementary conditions that ensure this behavior, we encountered that the noncommutative angular momentum term determined the relationship among the values of the scale factors of the cosmology at $\tau_B$. On the other hand, the criteria for a minimum resulted in a more complicated expression that in the general case is rather difficult to interpret in a geometrical sense, however, in the simple isotropic case, where the three scale factors take the same value $r_B$ at $\tau_B$, it greatly reduces to a more insightful form which imposes a lower bound for the value of $r_B$. It was observed that for the Planckian regime a consequence of this condition is that $r_B>1$, meaning that a bounce cannot occur for volume scales smaller than one Planck volume. It is also remarkable that this expression incorporates the contributions from quantum mechanics and noncommutativity in order for the minimum to acquire physically operational values. In this semiclassical context, proving the existence of bounce-like solutions implies that the probability of the quantum states to have singular behavior is marginal.

After implementing notions from solution methods frequently used in boundary-layer problems of fluid mechanics we inferred the asymptotic behavior of the analytical solution of the equations of motion in the vicinity of the bounce as well as far from it. This lead us to confirm the compatibility of the classical collapse and expansion solutions with the bounce mechanism induced by the new terms in the Hamiltonian. The linear approximation around the bounce indicates an interaction between the values of all the dynamical variables via the noncommutative corrections, which in the case where noncommutativity is not present decouples completely in three independent subspaces. It is worth mentioning that a missing piece of our analysis, due to the inherent technical complexities presented by the differential system, is the availability of an intermediate solution that bridges the outer solution with the inner solution, therefore we had to recur to computational resources to obtain a full picture of the evolution.

In the numerical results another aspect of the solutions was observed that is not evident from the asymptotic analysis, this corresponded to an apparent averaging of the scale factors around a mean trajectory when departing from anisotropic values at the bounce. This is suggestive to think that noncommutativity could act as a precursor of isotropy in macroscopic configurations, which from our numerical simulations appears to occur in the intermediate regime of the solutions.

\section*{Acknowledgments}

The authors are deeply grateful to Professor Marcos Rosenbaum for the enlightening discussions, profound suggestions and overall support during the preparation of this manuscript without which this research would have not been possible. Also to Professor A. Minzoni for taking his time providing useful insight on the behavior of the dynamical system and calling our attention to key aspects of the asymptotics.

% -----------------------------------------------------------
\bibliographystyle{ieeetr}
\bibliography{NCB}
\end{document}